\documentstyle[12pt,worldsci]{article}
\begin{document}

\title{{\bf PROBING NEW GAUGE BOSON COUPLINGS AT\\
HADRON SUPERCOLLIDERS}}
\author{THOMAS~G.~RIZZO\\
{\em High Energy Physics Division, Argonne National Laboratory , Argonne,\\
Illinois 60439, USA}\\
\vspace{0.3cm}
and\\
\vspace*{0.3cm}
{\em Ames Laboratory and Department of Physics, Iowa State University , Ames,\\
Iowa 50011, USA}}

\maketitle
\setlength{\baselineskip}{2.6ex}

\begin{center}
\parbox{13.0cm}
{\begin{center} ABSTRACT \end{center}
{\small \hspace*{0.3cm} Once it is discovered, the determination of the
various couplings of a new neutral gauge boson at a hadron supercollider will
not be
an easy task. We review several recent studies that have begun to examine this
issue for both the SSC and LHC.}}
\end{center}

\section{Introduction}

   If a new neutral gauge boson($Z_2$) is discovered at either the SSC or LHC
we will want to
know more about it than the fact that it exists. Since many models in the
literature predict such particles, we will want to know {\it which} $Z_2$ we
have discovered. To accomplish this goal one must probe the couplings of
the $Z_2$. This
will not be an easy task since only a few of the properties of the $Z_2$ are
easily measurable at hadron supercolliders. During the past 1-2 years there
have
been some initial attempts to address this problem and they will be reviewed
in the discussion below.

\section{Observables and Analyses}

      Neither the SSC or LHC detectors will have much difficulty in
making clean measurements of the mass($M_2$), width($\Gamma$), production
cross section($\sigma$), and forward-backward asymmetry($A_{FB}$) for a $Z_2$
in the TeV mass range{\cite 1} via the lepton-pair production channel. If this
is the {\it only} information available then identifying the $Z_2$ will be
difficult but not impossible {\it provided} that certain theoretical
prejudices are indeed realized in nature and sufficient statistics are
available. Within a given extended
gauge model, $\Gamma$ (and, hence $\sigma$) will be sensitive to what
final states are kinematically allowed in $Z_2$ decay since such models
require
the existence of exotic particles in addition to those present in the
Standard Model(SM).
{\it If} one assumes that the $Z_2$ can decay {\it only}
to the SM particles it is possible to get an excellent handle on the nature of
the $Z_2$ couplings using the above data alone{\cite 2}. However, such an
analysis could be spoiled by the potential contributions of these exotics
to $\Gamma$ or by the influence of a small mixing between the $Z_2$ and the SM
$Z$.
One can show that for a class of models, the probability that some of
these new degrees of freedom contribute substantially to $\Gamma$ is small
since it is very likely that they are more massive than $M_{2}/2$
and are thus kinematically forbidden to appear as decay products. Fig.~1a shows
this explicitly for the case of the exotic fermions in $E_6$ Effective Rank-5
Models{\cite 3}. Since the exotics and the $Z_2$ have their masses generated
by the same vev and all of the particle couplings are determined once the $E_6$
angle $-90^o\leq\theta\leq+90^o$ is fixed, the only free parameter for
each exotic particle is the size of its Yukawa coupling. This can be bounded
in the usual way by demanding tree-level unitarity in exotic particle
scattering amplitudes mediated by $Z_2$ exchange. One then can determine what
fraction of the allowed range for the Yukawa couplings will permit the exotics
to participate in $Z_2$ decay resulting in Fig.~1a; comparable bounds are
obtainable for the exotics in other models. Generally, however, the existence
of additional $Z_2$ decay modes remains a concern for analyses using only the
observables listed above.


\vspace*{5.9cm}
\noindent
{\small Fig. 1.  (a)Probability that a $Z_2$ can decay into the exotic fermions
 h,E,$S^c$(solid), $N$(dash-dot), or $N^c$(dash) in $E_6$ models. (b)$\chi^2$
fit
to the value of $\theta$ for the model discussed in the text; the solid(dashed)
curve corresponds to a $Z-Z_2$ mixing angle of 0.01(-0.01).}
\vspace*{0.6cm}

Similarly, it can also be shown that if the magnitude of the $Z-Z_2$ mixing
angle is small($\leq 0.01$, as might be expected for a heavy $Z_2$), then this
will have little effect on our ability to determine extended model couplings
using only the data above{\cite 3}; this is demonstrated by an explicit
example in
Fig.~1b for an $E_6$ model $Z_2$ with $\theta=0$(model $\psi$) and a $M_2$=3
TeV at the SSC for an integrated luminosity of 10$fb^{-1}$. Here, we perform a
$\chi^2$ fit to determine the value of $\theta$ obtainable from the above data
using
the properties of the SDC detector{\cite 1} {\it assuming} the absence of
mixing when it is in fact present and assuming decays to SM fermions only.
(We remind the reader that, once specified, the value of $\theta$ uniquely
fixes all of the $Z_2$ couplings.)
We see that the best-fit value of $\theta$
as well as its 95$\%$ CL allowed range(corresponding to the gap between the
curves along the dotted line) are only slightly altered by mixing
angles as large as 0.01{\cite 3}. We note that other models show more or less
the same sensitivity to finite mixing.

In order to circumvent the potential problems in coupling determinations
associated with theoretical uncertainties in the value of $\Gamma$ we must
search for new observables which are insensitive to this quantity. One
possibility is to examine 3-body decays such as
$Z_2{\rightarrow}W^{\pm}l^{\mp}\nu$ and $Z_2{\rightarrow}Zl^+l^-$ or $Z\nu\bar
\nu${\cite {4,5}}. For a relatively light $Z_2$ with a mass less than 1-2 TeV,
the number of events of this type is generally in the range $10^2-10^4$ at
both the SSC and LHC so that significant statistics can be accumulated once SM
backgrounds are removed{\cite 5}. The {\it ratios} of the number of these
kinds of
events to ordinary lepton-pair events is thus not too small and is $\Gamma$
independent. In particular, one defines{\cite 5} the ratios $r_{l\nu W}=
\Gamma(Z_2{\rightarrow}W^{\pm}l^{\mp}\nu)/\Gamma(Z_2{\rightarrow}l^+l^-)$ and
$r_{\nu\nu
Z}=\Gamma(Z_2{\rightarrow}Z\nu\bar\nu)/\Gamma(Z_2{\rightarrow}l^+l^-)
$ whose values are shown for a number of different models in Fig.~2a assuming
$M_2$=1 TeV and no $Z-Z_2$ mixing. (For a discussion of the individual
specific models shown, see Ref.~4.) Most models predict values for
these ratios that lie along the solid line {\it provided} the $Z_2$ has
generation-independent couplings and its corresponding generator commutes
with those of
$SU(2)_L$; this is indeed the case for most GUT-inspired models. By satisfying
these two conditions one finds that $r_{l\nu W}$ and $r_{\nu\nu Z}$ become
simultaneously correlated and bounded.
Scenarios {\it not}
satisfying these conditions will lie elsewhere on the plot.


\vspace*{5.9cm}
\noindent
{\small Fig. 2.  (a)$r_{l\nu W}$ and $r_{\nu\nu Z}$ for several different
extended gauge models as discussed in the text. (b)$r_{l\nu W}$ including
the effects of $Z-Z_2$ mixing as described in the text.}
\vspace*{0.6cm}

If mixing {\it does} occur these conditions are no longer satisfied as the
generator coupling to the $Z_2$ now has a small piece proportional to $T_{3L}$.
This does not significantly influence the resulting value of $r_{\nu\nu Z}$,
but $r_{l\nu W}$
can be greatly modified since mixing turns on an additional resonant diagram
involving the now non-zero $Z_2 W^+W^-$ coupling. This can result in a
substantial increase in the value of $r_{l\nu W}$ as shown in Fig.~2b for the
$E_6$ model case as a function of $\theta$ (x-axis) and the ratio of the two
Higgs-doublets vevs, tan$\beta$ (y-axis) assuming $M_2$=1 TeV. Here we see that
mixing can induce values for $r_{l\nu W}$ of order unity or larger for a
broad region of parameter space. The result of mixing for a model lying
along the solid curve in Fig.~2a would then be a shift to the right without
any appreciable shift up or down. While discovering a $Z_2$ whose values of
these ratios place it in the lower right-hand part of Fig.~2a would be
difficult to interpret (something exotic and/or non-zero mixing), a value of
$r_{\nu\nu Z}\geq$0.06 would be a clear indication that something unusual has
been found. (Unfortunately, $r_{\nu\nu Z}$ has a very serious SM background
from $2Z$ production that is very difficult to deal with{\cite 5}.) A study of
the SM backgrounds for the above 3-body channels, including the decays of the
final state $W$ and $Z$, has recently been done by del Aguila et al.{\cite 6}
for a number of different extended models. These authors conclude that the
$Z_2{\rightarrow}e\bar\mu$ plus missing energy final states are reasonably
sensitive to $Z_2$ couplings and are statistically powerful provided that
$M_2$ is less than about 1.5 TeV.

As a last point we mention that in some extended models the $Z_2$ is
accompanied by a $W_2$ with a comparable mass; in a large fraction of cases
the two particles are almost exactly degenerate so that $W_2$ cannot
participate in $Z_2$ decay. (Of course, the mere observation of a $W_2$ will
tell us a great deal about the nature of the extended model and in most cases
the $W_2$ production cross section is larger than that of the $Z_2$ making it
likely that $W_2$'s might be observed first as was the case for the SM gauge
bosons.) In the Left-Right Symmetric Model(LRM) however, there is a
region of parameter space which allows $Z_2{\rightarrow}W_2^+W_2^-$ the rate
for which depends on the nature of $SU(2)_R$ breaking and the ratio of the
$SU(2)_L$ and $SU(2)_R$ coupling constants, $\kappa$. A somewhat larger range
of parameters allows for the corresponding 3-body decay $Z_2{\rightarrow}W_2
^{\pm}l^{\mp}\nu$. An observation of either of these modes will provide us
much needed information on the extended gauge sector. For a further discussion
of these possibilities see Refs.~4 and 7.
%

\vspace*{5.9cm}
\noindent
{\small Fig. 3.  (a)$\tau$ polarization asymmetry, $A$,  as a function of
$\theta$ for the case discussed in the text. (b)Left-right asymmetry, $A_{LR}$,
 at the SSC for both the $E_6$ model with $M_2$=1(solid) or 2(dotted) TeV  and
the
LRM for $M_2$=1(dashed) or 2(dot-dashed) TeV as discussed in the text.}
\vspace*{0.6cm}

Another possibility, which has a long history and has been recently resurrected
 {\cite 8}, is to measure the polarization asymmetry, $A$, of taus coming from
the
decay $Z_2{\rightarrow}\tau^+\tau^-$.
The advantage of this observable is that, in the
narrow-width approximation, it {\it directly} probes the leptonic $Z_2$
couplings
and is very insensitive to structure function and luminosity uncertainties:
$A=-2v_\tau a_\tau/(v_\tau^2+a_\tau^2)$. Fig.~3a
shows the strong sensitivity of $A$ to the mixing parameter $\theta$ discussed
above for $E_6$ models.
Unfortunately, observing $\tau$ pairs at hadron supercolliders
is difficult due to substantial backgrounds from $t\bar t$ and $W^+W^-$
pairs as well as conventional QCD 2-jet production all of which must be
drastically reduced before
the value of $A$ can be reliably determined; this problem has been recently
addressed for the SDC by Anderson et al{\cite 8}. These authors have
shown (for a $Z_2$ arising from the $E_6$ model discussed above with $M_2$=1
TeV) that a judicious choice of cuts can reject as much as 97$\%$ of the
background  and provides a determination of $A$ at the 10$\%$ level assuming a
luminosity of 10 $fb^{-1}$. (After these cuts only 6$\%$ of the $\tau$-pairs
from $Z_2$ decay remain but, since event rates are high, enough statistics
remains available.) Unfortunately, for a heavier $Z_2$ it is unlikely
that $\tau$ polarization data will be very useful as the signal to background
ratio is substantially smaller even if larger luminosities were available.
These authors{\cite 8} hypothesize that a better choice of cuts may help this
situation; clearly more work on this observable is needed.

If {\it polarized} proton beams become available at either the SSC or LHC, then
 other asymmetries such as the
left-right asymmetry, $A_{LR}$ (defined in a manner similar to what is
discussed for $e^+e^-$ collisions), can be constructed{\cite 9} which are
comparable in magnitude to the more conventional $A_{FB}$. Fiandrino and Taxil
{\cite 9}, have recently shown that such asymmetries are quite sensitive to
the choice of extended model as well the values of model parameters, such as
$\theta$ in the $E_6$ scenario, as shown in Fig.~3b. (In this plot, $\beta$ =
$90^o-\theta$ and $\alpha$ is related to the parameter $\kappa$ of the LRM.)
Unfortunately, even
if such asymmetries were reliably determined for the dilepton invariant mass
region near $M_2$ they would be difficult to interpret in terms of model
couplings in a more than a qualitative fashion due to the very large
uncertainties currently present in the polarized parton densities.  The
authors of Ref.~8 are optimistic, however, that new polarized scattering data
anticipated from RHIC may alleviate at least some of these difficulties.

\section{Conclusions}

As can be seen from the discussion above, the identification of new $Z_2$ gauge
 bosons at hadron supercolliders remains a serious problem especially if the
mass of this particle exceeds 2 TeV. Clearly, much more work will be needed to
address the issues raised here before hadron supercolliders are turned on later
 in the decade.

\bibliographystyle{unsrt}

\end{document}